\documentclass[prl,twocolumn,showpacs,aps]{revtex4-1}

\usepackage{amsmath, amsfonts, amssymb, mathrsfs}
\usepackage{txfonts}
\usepackage{graphicx}
\usepackage{epsfig}
\usepackage{enumerate}
\usepackage{caption}
\usepackage{subcaption}

\usepackage{slashed}
\usepackage{ulem}

\newcommand{\bea}{\begin{eqnarray}}
\newcommand{\eea}{\end{eqnarray}}

\newcommand{\be}{\begin{equation}}
\newcommand{\ee}{\end{equation}}

\usepackage{ifpdf}
\ifpdf
\usepackage{epstopdf}
\fi

\usepackage[unicode]{hyperref}
\hypersetup{
	pdftitle={Replica Phase Transition with Quantum Gravity Corrections},
	pdfauthor={Nian, Zhong},
	pdfsubject={Replica Phase Transition with Quantum Gravity Corrections},
	colorlinks=true,
	linkcolor=blue,
	citecolor=blue,
	filecolor=black,
	urlcolor=blue
}

\def\url#1{}

\makeatletter
\newcommand{\vast}{\bBigg@{4}}
\newcommand{\Vast}{\bBigg@{5}}
\makeatother

\usepackage{bm}
\usepackage{color}

\begin{document}

\title{Replica Phase Transition with Quantum Gravity Corrections}
\author{Jun Nian$^{1}$}

\author{Yuan Zhong$^{1, 2}$}
\email{zhongyuan@ucas.ac.cn}

\affiliation{$^1$ International Centre for Theoretical Physics Asia-Pacific, University of Chinese Academy of Sciences, 100190 Beijing, China}
\affiliation{$^2$ The School of Science, Great Bay University, Dongguan, Guangdong, 523000, China}

\begin{abstract}
Motivated by bulk replica wormholes, we study the boundary effective theory that describes the near-horizon fluctuations of a near-extremal Reissner-Nordstr\"om black hole. This theory consists of a Schwarzian mode and a $U(1)$ phase mode. We compute the partition function of this boundary theory on connected geometries, from which the entropy is derived. Our analysis reveals a rich phase structure, in which the dominance of connected or disconnected configurations leads to a phase transition controlled by the temperature and the coupling constants $C$, $K$, and $\mathcal{E}$ of the 1d effective theory.
\end{abstract}

\maketitle

\textit{Introduction.---} The study of black hole thermodynamics has revealed that the low-energy dynamics of near-horizon near-extremal Reissner-Nordstr\"om (RN) black holes is governed by the AdS$_2$ JT gravity coupled to a Maxwell theory describing the near-horizon fluctuations \cite{Cadoni:2008mw, Almheiri:2016fws, Moitra:2018jqs, Iliesiu:2019lfc, Iliesiu:2020qvm}. This 2d theory, which is dual to a 1d boundary effective theory \cite{Sachdev:2019bjn, Iliesiu:2020qvm, Davison:2016ngz, Gaikwad:2018dfc} consisting of a Schwarzian mode (encoding gravitational degrees of freedom) and a $U(1)$ phase mode (encoding electromagnetic fluctuations), has become a pivotal model for understanding quantum gravity in low dimensions. Our research is motivated by the 2d bulk replica wormhole mechanism \cite{Penington:2019kki}, but is executed entirely within this 1d boundary effective theory. We compute the partition function of the coupled Schwarzian and $U(1)$ theory on replica geometries, thereby deriving the entropy from the boundary perspective. This analysis uncovers a rich phase structure, with transitions between connected and disconnected saddles controlled by temperature and effective theory coupling constants $C$, $K$, and $\mathcal{E}$. 

The novel feature of our work is that, compared to known replica wormhole-sourced phase transitions between connected and disconnected replica geometries as $n\to1$, we do not require any additional outer entangled system, such as a thermal bath or end-of-world branes \cite{Almheiri:2019qdq, Penington:2019kki, Goto:2020wnk, Marolf:2020xie}. Instead, we study a system only consisting of Schwarzian modes and $U(1)$ gauge modes. The phase transition originates from the competition among coupling constants.

\textit{Effective Theory of Near-Horizon Near-Extremal RN Black Hole.---} In this section, we quickly review the effective action describing the near-horizon fluctuations of a near-extremal AdS$_{d+2}$ RN black hole \cite{Sachdev:2019bjn}, i.e., the 2d JT gravity coupled to the Maxwell theory. This 2d theory is dual to the boundary 1d effective theory
\begin{align}\label{action_Sch_U(1)}
  I_{\text{eff}} & = -S_0 -C \int_0^\beta d\tau \{\tan \pi T f(\tau),\tau\} \nonumber\\
  {} & \quad +\frac{K}{2} \int_0^\beta d\tau \left( \partial_\tau\phi -i(2\pi T{\mathcal{E}})\partial_\tau f\right)^2 \, ,
\end{align}
where the effective theory couplings $C$, $K$, $\mathcal{E}$ obtained from dimensional reduction are related to the higher-dimensional couplings as
\be\label{def:C}
C =\frac{s_d \Phi_1}{\kappa^2}=\frac{d s_d R_h^{d-1}R_2^2}{\kappa^2}\, ,
\ee
\be\label{def:K}
K=\frac{ (d-1)s_d R_h^{d-3} [ d(d+1)R_h^2 +(d-1)^2 L^2 ] }{(d+1)g_F^2}\, ,
\ee
\be\label{def:calE}
{\mathcal{E}} = \frac{g_F R_h L \sqrt{d\left[(d+1)R_h^2 + (d-1)L^2\right]}}{\kappa \left[d(d+1)R_h^2 + (d-1)^2 L^2\right]}\, ,
\ee
with
\be
R_2 = \frac{L R_h}{\sqrt{d(d+1)R_h^2 + (d-1)^2 L^2}}
\ee
denoting the AdS$_2$ radius and $L$ standing for the AdS$_{d+2}$ radius. Here, $R_h$ is the horizon radius of the extremal AdS$_{d+2}$ RN black hole. This theory has four independent parameters $(C, K, {\mathcal{E}}, \beta)$, or equivalently, $(\kappa, g_F, \mu_0, \beta)$, where $\kappa$, $g_F$, $\mu_0$, $\beta$ denote the gravitational constant, the Yang-Mills coupling constant, the chemical potential, and the inverse temperature, respectively.

At finite temperature, the AdS$_{d+2}$ RN black hole's entropy is
\begin{equation}
  S(T,\mu) = \frac{2\pi s_d}{\kappa^2} r_0(T,\mu)^d\, ,
\end{equation}
where $r_0(T,\mu)$ is the horizon radius of the AdS$_{d+2}$ RN black hole at finite temperature. When $T=0$, the entropy becomes
\begin{align}
S_0 \equiv S(T=0,\mu_0) & = \frac{2\pi s_d}{\kappa^2} R_h^d\, , \label{S_0(R_h)}
\end{align}
where the extremal black hole entropy $S_0$ depends on the parameters $C$, $K$, and ${\mathcal{E}}$.

Using \eqref{def:C}, \eqref{def:K} and \eqref{def:calE}, we can express $\kappa$, $g_F$, and $R_h$ in terms of $C$, $K$ and $\mathcal{E}$. Substituting the expression of $R_h$ back into \eqref{S_0(R_h)}, we can express $S_0$ in terms of the boundary theory parameters. This is doable, but the result is rather complicated for AdS$_{d+2}$ with general dimensions. From now on, we focus on the special case AdS$_4$ with $d=2$. In this case, the zero-temperature entropy $S_0$ reads
\be
S_0^{d=2}=\frac{\sqrt{3} \pi \sqrt{C} (\mathcal{E}^2 K+C)}{L \sqrt{\mathcal{E}^2 K - C}}\, ,
\ee
which is real and positive, provided
\be\label{lowerbound:E}
  {\mathcal{E}}^2 > C/K.
\ee
If the lower bound \eqref{lowerbound:E} on $\mathcal{E}$ is violated, the entropy $S_0$ will become imaginary. This suggests that there could be a quantum phase transition of the topological term $S_0$ in the boundary theory \eqref{action_Sch_U(1)} at ${\mathcal{E}}^2 = C/K$. This phase transition appears to be between AdS$_2$ JT gravity and dS$_2$ JT gravity. They both admit the Schwarzian effective theory descriptions, but the $S_0$ term is real for AdS while imaginary for dS \cite{Cotler:2019nbi, Maldacena:2019cbz}.

A nice feature of the effective theory is that the phase mode and the Schwarzian mode decouple in the partition function \cite{Mertens:2019tcm}:
\be
Z = Z_{\text{SL}(2)}[C,\beta] \,\, Z_{U(1)}[K,{\mathcal{E}},\beta]\, ,
\ee
where ${\mathcal{E}}$ only appears only in the phase-mode part $Z_{U(1)}$ through the chemical potential as
\be\label{ZU1-bdy}
Z_{U(1)} = \theta_3 \left( e^{-\beta/K},\, e^{-2\pi {\mathcal{E}}} \right)\, .
\ee
When $\beta/K\gg 1$, or equivalently $q=e^{-\beta/K} \to 0$, the function $\theta_3$ has the expansion: $\theta_3(q,e^{2\pi i z}) =1 + 2 q \cos(2z) + O(q^2)$. Therefore,
\be
Z_{U(1)} =  1+ 2e^{-\beta/K} \cosh(2 {\mathcal{E}}) +O(e^{- 2 \beta/K}).
\ee

\textit{Partition Function of Effective Theory on the Replica Geometry as $n\to 1$.---} Our goal is to provide the replica partition function for the boundary effective theory \eqref{action_Sch_U(1)} in the limit $n\to 1$. For the Schwarzian part, this procedure has already been done in \cite{Penington:2019kki}. The $n\to 1$ on-shell action for the Schwarzian field on the quotient manifold $M_n/Z_n$ is
\be\label{action_sch_Mn}
I^{M_n/Z_n}_{\text{Sch}} \bigg|_{n\to1}= -\frac{2\pi^2 C}{\beta} -\frac{8\pi^2 C}{\beta} \cosh\rho \,(n-1) +O((n-1)^2)\, ,
\ee
where $2\rho$ is the geodesic distance between the two $Z_n$-fixed points. We quickly review the calculation of the Schwarzian action on the quotient geometry $M_n/Z_n$, which will be discussed in more detail later. The quotient geometry $M_n/Z_n$ can be regarded as the trumpet geometry \eqref{trumpet_metric}, where the trumpet parameter $b$ is related to the distance $\rho$ between the $Z_n$ fixed points as \eqref{b_rho} when $n\to 1$. Then the Schwarzian action $I^{M_n/Z_n}_{\text{Sch}}$ on the quotient geometry can be computed as the Schwarzian action on the trumpet geometry as $n\to1$
\be
I^{M_n/Z_n}_{\text{Sch}} \bigg|_{n\to1} = \frac{2\pi^2 C}{\beta}\left[1+ \frac{b^2}{2\pi^2}\right].
\ee
Substituting the relation \eqref{b_rho} between $b$ and $\rho$ and expanding the action to the order of $(n-1)$, we obtain the Schwarzian action \eqref{action_sch_Mn}.

The action on the original replica geometry $M_n$ before the $Z_n$ quotient as $n\to1$ is
\begin{align}
I^{n\to 1}_{\text{Sch}} & = \left[n I^{M_n/Z_n}_{\text{Sch}}\right]_{n \to 1}\nonumber \\
  & = -\frac{2\pi^2 C}{\beta} -\frac{2\pi^2 C}{\beta}(1+4 \cosh\rho) \,(n-1) +O((n-1)^2)\, .
\end{align}

The remaining task is to do a similar calculation for the $U(1)$ phase mode. To achieve this goal, we use the following trick. We first calculate the bulk $U(1)$ partition function on the replica geometry and find its dependence on the replica parameter $n$. With the assumption that the bulk partition function equals the boundary partition function, we can replace the 2d YM parameters by the boundary effective theory parameters to obtain the boundary $U(1)$ partition on the replica geometry.

Let us consider an $n$-replica geometry. For the 2d Maxwell theory, the partition function on this $n$-hole surface can be obtained from the standard 2d YM method by summing over representations of $U(1)$, which equals a sum over integers \cite{Cordes:1994fc, Blommaert:2018oue}.

\begin{figure}[!htb]
  \begin{center}
    \includegraphics[width=0.5\textwidth]{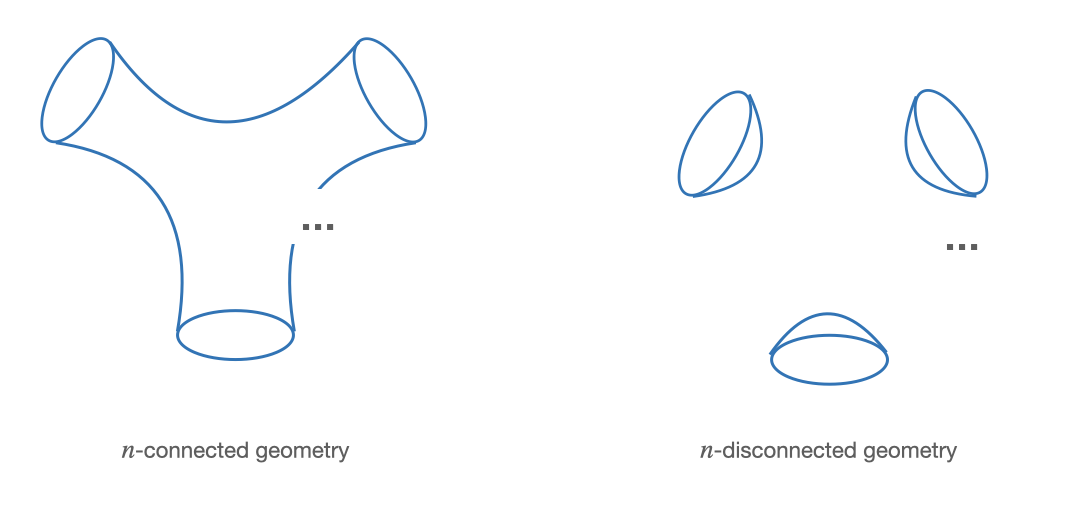}
    \caption{The connected and disconnected replica geometries}\label{fig:replica}
  \end{center}
\end{figure}

For the disconnected $n$-replica geometry, the topology is just $n$ copies of caps (see Fig.~\ref{fig:replica}). The partition function on this geometry is the $n$-th power of the partition function on a single cap with chemical potential $\mu$, which is related to the gauge field $A_\tau\big|_{\partial}=i\mu$. The corresponding holonomy around the boundary thermal circle is $U=e^{\mu\beta}$. Hence,
\be
  \left[Z(a_1,U)\right]^{ n} = \left[\sum_{q\in \mathbb{Z}} e^{\mu\beta q} e^{-\frac{e^2}{2} a_1 q^2}\right]^n = \theta_3\left(e^{-{e^2 a_1}},\, e^{-\mu\beta}\right)^n\, ,
\ee
where $a_1(\beta)$ is the area of a hyperbolic cap with the thermal circle period $\beta$.

For the connected $n$-replica wormhole, the topology is a hyperbolic surface with $n$ circle boundaries with the asymptotic length $\beta$ for each (see Fig.~\ref{fig:replica}). The $U(1)$ partition function for this case is
\be
  Z[a_n,U,\cdots,U] = \sum_q \left(e^{\mu\beta q}\right)^n \, e^{-\frac{e^2}{2}a_n q^2} = \theta_3\left(e^{-{e^2 a_n}},e^{-n \mu\beta}\right)\, ,
\ee
where $a_n(\beta)$ is the area of the $n$-connected surface described above. Comparing these expressions with the boundary $U(1)$ partition function \eqref{ZU1-bdy}, we infer the replica $U(1)$ partition function in terms of boundary effective theory parameters as
\be
Z_{U(1)}^{M_n} = \theta_3 \left( e^{-\frac{\beta}{K} \frac{a_n}{a_1}} , e^{-2\pi n {\mathcal{E}}} \right)\, .
\ee

Assuming that the bulk partition function equals the boundary effective theory partition function, we obtain the boundary phase mode's partition function from the bulk Maxwell partition function in terms of boundary quantities. In addition, to calculate the entropy, we also employ the replica trick and expand the partition function in the limit $n\to 1$.
\begin{align}
  Z^{n\to 1}_{U(1)} & = \theta_3 \left( e^{-\frac{\beta}{K} \frac{a_n}{a_1}}, e^{-2\pi n {\mathcal{E}}} \right) \nonumber\\
  {} & = \theta_3 \left( e^{-\frac{\beta}{K}}, e^{-2\pi  {\mathcal{E}}} \right) + \Bigg[ \partial_q \theta_3(q,\eta)\, e^{-\frac{\beta}{K}} \left(-\frac{\beta}{K}\right)\frac{a_n-a_1}{a_1} \nonumber\\
  {} & \qquad + \partial_\eta \theta_3(q,\eta)\, e^{-2\pi  {\mathcal{E}}}\, (-2\pi  {\mathcal{E}})\, (n-1) \Bigg] + O((n-1)^2)\, ,
\end{align}
where the differentials are evaluated at $q=e^{-\beta/K}$, $\eta=e^{-2\pi{\mathcal{E}}}$.

\begin{figure}[!htb]
  \begin{center}
    \includegraphics[width=0.45\textwidth]{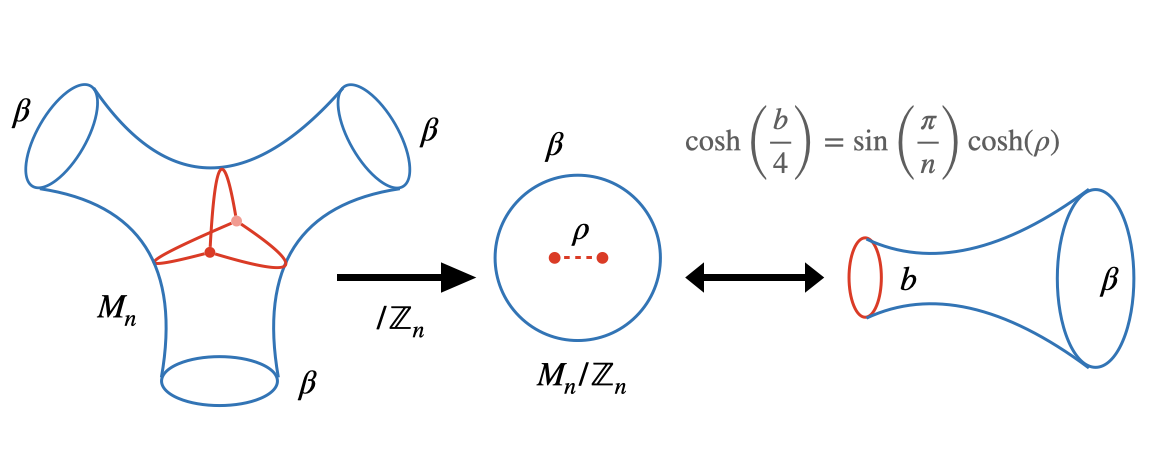}
    \caption{The quotient geometry $M_n/Z_n$ can be treated as a trumpet geometry}\label{fig:trumpet}
  \end{center}
\end{figure}

To determine the ratio of areas $a_n$ over $a_{1}$. Recall that the quotient geometry $M_n/Z_n$ can be approximated by the trumpet geometry \cite{Penington:2019kki} (see Fig.~\ref{fig:trumpet}):
\be\label{trumpet_metric}
ds^2=d\sigma^2 + \cosh^2{\sigma}\, \frac{b^2}{(2\pi)^2}\, d\theta^2\, ,
\ee
where the trumpet parameter $b$ is related to the geodesic distance $\rho$ between fixed points as
\be
  \cosh \left(\frac{b}{4}\right) = \sin \left(\frac{\pi}{n}\right)\, \cosh(\rho)\, ,
\ee
and in the limit $n\to 1$
\be\label{b_rho}
  b = 2\pi i -4\pi i \cosh(\rho)\, (n-1) + O((n-1)^2)\, .
\ee

For the ratio of areas, $a(M_n)/a(M_1)$, near $n=1$, note that

\be
a(M_n)=n \, a(M_n/Z_n) =n \int_0^{2\pi} d\theta \int_0^\infty d\sigma \cosh(\sigma) \frac{b_n}{2\pi} \propto nb_n\, .
\ee
Hence, the ratio of areas in the limit $n\to 1$ is
\begin{align}
  {} & \frac{a_n-a_1}{a_1} =\frac{n b_n-b_1}{b_1} \nonumber\\
  =\, & \frac{n( 2\pi i -4\pi i \cosh(\rho)(n-1) +O((n-1)^2) -2\pi i }{2\pi i }\nonumber\\
  =\, & (1-2\cosh\rho)(n-1) +O((n-1)^2)\, .
\end{align}
Consequently, the $U(1)$ partition function near $n=1$ is
\begin{align}
  Z^{n\to 1}_{U(1)} =\, & \theta_3 \left( e^{-\frac{\beta}{K}}, e^{-2\pi  {\mathcal{E}}} \right) \nonumber\\
  {} & + (n-1) \Bigg[ \partial_q\theta_3(q,\eta)\, e^{-\frac{\beta}{K}} \left(-\frac{\beta}{K}\right)\, (1-2\cosh(\rho))\nonumber\\
  {} & \qquad\qquad + \partial_\eta\theta_3(q,\eta)\, e^{-2\pi {\mathcal{E}}}\, (-2\pi  {\mathcal{E}})\Bigg] \nonumber\\
  {} & + O((n-1)^2)\, ,
\end{align}
whose logarithm gives the free energy
\begin{align}
  I^{n\to 1}_{U(1)} & = -\log Z^{n\to 1}_{U(1)} \nonumber\\
  {} & = -\log \theta_3 \left( e^{-\frac{\beta}{K}}, e^{-2\pi  {\mathcal{E}}} \right) +  (n-1) I^{(1)}_{U(1)} + O((n-1)^2)\, ,
\end{align}
where the order $(n-1)$ part of the free energy reads
\begin{align}
I^{(1)}_{U(1)} =\, & -\theta_3 \left( e^{-\frac{\beta}{K}}, e^{-2\pi  {\mathcal{E}}} \right)^{-1} \nonumber\\
{} & \cdot \Bigg[ \partial_q\theta_3(q,\eta) e^{-\frac{\beta}{K}} (-\frac{\beta}{K})(1-2\cosh(\rho))\nonumber\\
{} & \qquad + \partial_\eta\theta_3(q,\eta) e^{-2\pi  {\mathcal{E}}} (-2\pi  {\mathcal{E}})\Bigg]\, .
\end{align}
In the low-temperature limit, the total free energy (besides the topological term $-\chi S_0$) is
\begin{align}
  {} & I^{n\to 1}_{\text{tot}} = I_{\text{Sch}} + I_{U(1)} \nonumber\\
  =\, & - \frac{2\pi^2 C}{\beta} -\frac{2\pi^2 C}{\beta}(1+4 \cosh\rho) \,(n-1) -\log\theta_3 \nonumber\\
  {} & - (n-1) \left[\theta_3^{-1} \partial_q\theta_3(q,\eta) e^{-\frac{\beta}{K}} \frac{-\beta}{K}(1-2\cosh\rho) - \theta_3^{-1}\partial_\eta\theta_3 e^{-2\pi{\mathcal{E}}}2\pi{\mathcal{E}} \right] \nonumber\\
  {} & + O(n-1)^2\, ,
\end{align}
which reaches extremality at $\rho=0$.

\textit{Entropy of Effective Theory.---} The total ($n\to 1$) connected geometry boundary action consists of the Schwarzian part, the $U(1)$ part, and the topological part.
\begin{align}
{} & \log Z_n^{\text{conn}} \sim \chi S_0 -I^{n\to 1}_{\text{Sch}} -I^{n\to 1}_{U(1)} \nonumber\\
=\, & (2-n)S_0 +\frac{2\pi^2 C}{\beta} +\log\theta_3 \nonumber\\
{} & + (n-1) \left[ \frac{10\pi^2 C}{\beta} +\theta_3^{-1}\partial_q\theta_3 e^{-\frac{\beta}{K}} \frac{\beta}{K}+\theta_3^{-1}\partial_\eta\theta_3 e^{-2\pi  {\mathcal{E}}} (-2\pi  {\mathcal{E}})\right] \nonumber\\
{} & + O((n-1^2))\, .
\end{align}

We can apply the replica trick to calculate the entropy of the system and its copies sourced by the connected geometry
\be
  S = \lim_{n\to 1} \frac{\log Z_n -n\log Z_1}{n-1} = \partial_{n=1} \log Z_n -\log Z_1\, .
\ee
The connected entropy can be recomposed as $S^{\text{conn}} = \lim_{n\to 1_+} \frac{\log Z^{\text{conn}}_n -\log Z_1^n}{n-1} =  \lim_{n\to 1_+} \frac{\log Z^{\text{conn}}_n -\log Z^{\text{disconn}}_n}{n-1}$. When $S^{\text{conn}}>0$, $ -\log Z^{\text{conn}}_n<-\log Z^{\text{disconn}}_n$ for $n\approx 1_+$, thus the disconnected phase has larger free energy and dominates, and vice versa. This agrees with the following discussions comparing entropy.

For $\log Z_n^{\text{conn}}$, we get the connected-geometry-sourced entropy
\begin{align}
S^{\text{conn}} & = -S_0 +\left[ \frac{10\pi^2 C}{\beta} +\theta_3^{-1}\partial_q\theta_3 e^{-\frac{\beta}{K}} \frac{\beta}{K}-\theta_3^{-1}\partial_\eta\theta_3 e^{-2\pi  {\mathcal{E}}} 2\pi  {\mathcal{E}}\right] \nonumber\\
{} & \quad -\left( S_0 +\frac{2\pi^2 C}{\beta} +\log\theta_3 \right) \nonumber\\
{} & = -2S_0 +\frac{8\pi^2 C}{\beta} -\log\theta_3 +\theta_3^{-1}\partial_q\theta_3 e^{-\frac{\beta}{K}} \frac{\beta}{K}\nonumber\\
{} & \quad - \theta_3^{-1}\partial_\eta\theta_3 e^{-2\pi{\mathcal{E}}}2\pi{\mathcal{E}}\, .
\end{align}
For the disconnected geometry,
\be
S^{\text{disconn}} = 0\, ,
\ee
which always vanishes since $\log Z^{\text{disconn}}_n = \log Z_1^n \propto n$.

We can distinguish different cases. If 
\be
-2S_0 +\frac{8\pi^2 C}{\beta} -\log\theta_3 +\theta_3^{-1}\partial_q\theta_3 e^{-\frac{\beta}{K}} \frac{\beta}{K}  -\theta_3^{-1}\partial_\eta\theta_3 e^{-2\pi{\mathcal{E}}}2\pi{\mathcal{E}}>0\, ,
\ee
then $S^{\text{conn}}>0$, and the von Neumann entropy of the system is governed by the disconnected one, i.e., $S=S^{\text{disconn}}=0$. If
\be
-2S_0 +\frac{8\pi^2 C}{\beta} -\log\theta_3 +\theta_3^{-1}\partial_q\theta_3 e^{-\frac{\beta}{K}} \frac{\beta}{K}  -\theta_3^{-1}\partial_\eta\theta_3 e^{-2\pi{\mathcal{E}}}2\pi{\mathcal{E}}<0\, ,
\ee
then $S^{\text{conn}}<0$, and the von Neumann entropy of the system is governed by the connected one, i.e., $S=S^{\text{conn}}<0$.

In the low-temperature limit, $\beta/K \gg 1$, we can expand the $\theta_3$ function in $Z_{U(1)}$ and obtain
\begin{align}
\log Z_{U(1)} & = \log\theta_3(e^{-\beta/K},e^{-2\pi{\mathcal{E}}}2\pi{\mathcal{E}}) \nonumber\\
{} & = 2e^{-\beta/K} \cosh(2 {\mathcal{E}}) +O(e^{-\beta/K})^2\, .
\end{align}
Consequently, the connected entropy reads
\begin{align}
{} & S^{\text{conn}} \nonumber\\
=\, & -2S_0 +\frac{8\pi^2 C}{\beta} -\log\theta_3 +\partial_q\log\theta_3 e^{-\frac{\beta}{K}} \frac{\beta}{K}  -\partial_\eta\log\theta_3 e^{-2\pi{\mathcal{E}}}2\pi{\mathcal{E}} \nonumber\\
\approx\, & -2S_0 +\frac{8\pi^2 C/K}{\beta/K} +2\cosh(2 {\mathcal{E}}) e^{-\frac{\beta}{K}}\frac{\beta}{K} +4e^{-\frac{\beta}{K}}  {\mathcal{E}}\sinh(2 {\mathcal{E}})\, ,
\end{align}
where we neglect higher-order terms, and
\be
S_0=\frac{\sqrt{3} \pi \sqrt{CK} (\mathcal{E}^2 +C/K)}{L \sqrt{\mathcal{E}^2 - C/K}}\, .
\ee

\textit{Phase Transitions.---} Now, we are ready to discuss the phase structure between the connected and the disconnected phases.

\paragraph*{i) Transition on $\beta$.}

For fixed $C$, $K$, and ${\mathcal{E}}^2>C/K$, the terms in $S^{\text{conn}}$ other than $-2 S_0$ is
\begin{align}
S^{\text{conn}}_* & \equiv S^{\text{conn}} +2S_0 \nonumber\\
{} & = \frac{8\pi^2 C/K}{\beta/K} +2\cosh(2 {\mathcal{E}}) e^{-\frac{\beta}{K}}\frac{\beta}{K} +4e^{-\frac{\beta}{K}}  {\mathcal{E}}\sinh(2 {\mathcal{E}})\nonumber\\
{} & \quad + O((e^{-\beta/K})^2)\, ,
\end{align}
which monotonically decays when $\beta/K$ increases. The transition point is when $S^{\text{conn}}$ crosses $0$, or equivalently, when $S^{\text{conn}}_*$ crosses $2S_0$. When $\beta$ grows from $0$ to $\infty$, $S^{\text{conn}}_*$ decays from $\infty$ to $0$, which always intersects with $2S_0$ at some finite $\beta$. Therefore, in the relatively high temperature region (small $\beta$), $S^{\text{conn}} >0$, the disconnected phase $S^{\text{disconn}}=0$ dominates; in the relatively low temperature region, $S^{\text{conn}} <0$, the connected phase dominates.

\paragraph*{ii) Transition on $\mathcal{E}$.}

When we fix $C$ and $K$, and increase $\mathcal{E}$ from $\sqrt{C/K}$ to $\sqrt{3C/K}$, $S_0$ first decays from $\infty$ to
\be
S_0^{\text{min}} = \frac{2\sqrt{6} \pi C }{L},\, \text{ at } \mathcal{E}=\sqrt{3C/K}\, ;
\ee
then as $\mathcal{E}$ goes from from $\sqrt{3C/K}$ to $\infty$, $S_0$ increases from $S_0^{\text{min}}$ to $\infty$, and grows linearly on $\mathcal{E}$ when $\mathcal{E} \gg \sqrt{C/K}$.

While the term $S^{\text{conn}}_* $ monotonically grows with $\mathcal{E}$. That means when $\mathcal{E}=\sqrt{C/K}$, $S^{\text{conn}}_* \big|_{{\mathcal{E}=\sqrt{C/K}}} < \infty$ smaller than $2S_0$; when $\mathcal{E} \gg \sqrt{C/K}$, $S^{\text{conn}}_* $ grows exponentially in $\mathcal{E}$ and larger than $2S_0$, which grows linearly with $\mathcal{E}$. Hence, when $\mathcal{E}>\sqrt{C/K}$ small, $S^{\text{conn}}<0$ dominates; when $\mathcal{E}$ large and $S^{\text{conn}}>0$, $S^{\text{disconn}}$ dominates. We conclude that there must be at least one connected-disconnected phase transition on $\mathcal{E}$ from $\sqrt{C/K}$ to $\infty$.

\paragraph*{iii) Transition on $C/K$ with $C/K \in (0, \mathcal{E}^2)$.}

When $\beta/K$ is relatively large, $S^{\text{conn}}_*$ is suppressed and always smaller than $2S_0$. In this case, the connected phase always dominates, and there is no phase transition on $C/K$.

\begin{figure}[!bth]
      \begin{center}
        \includegraphics[width=0.314\textwidth]{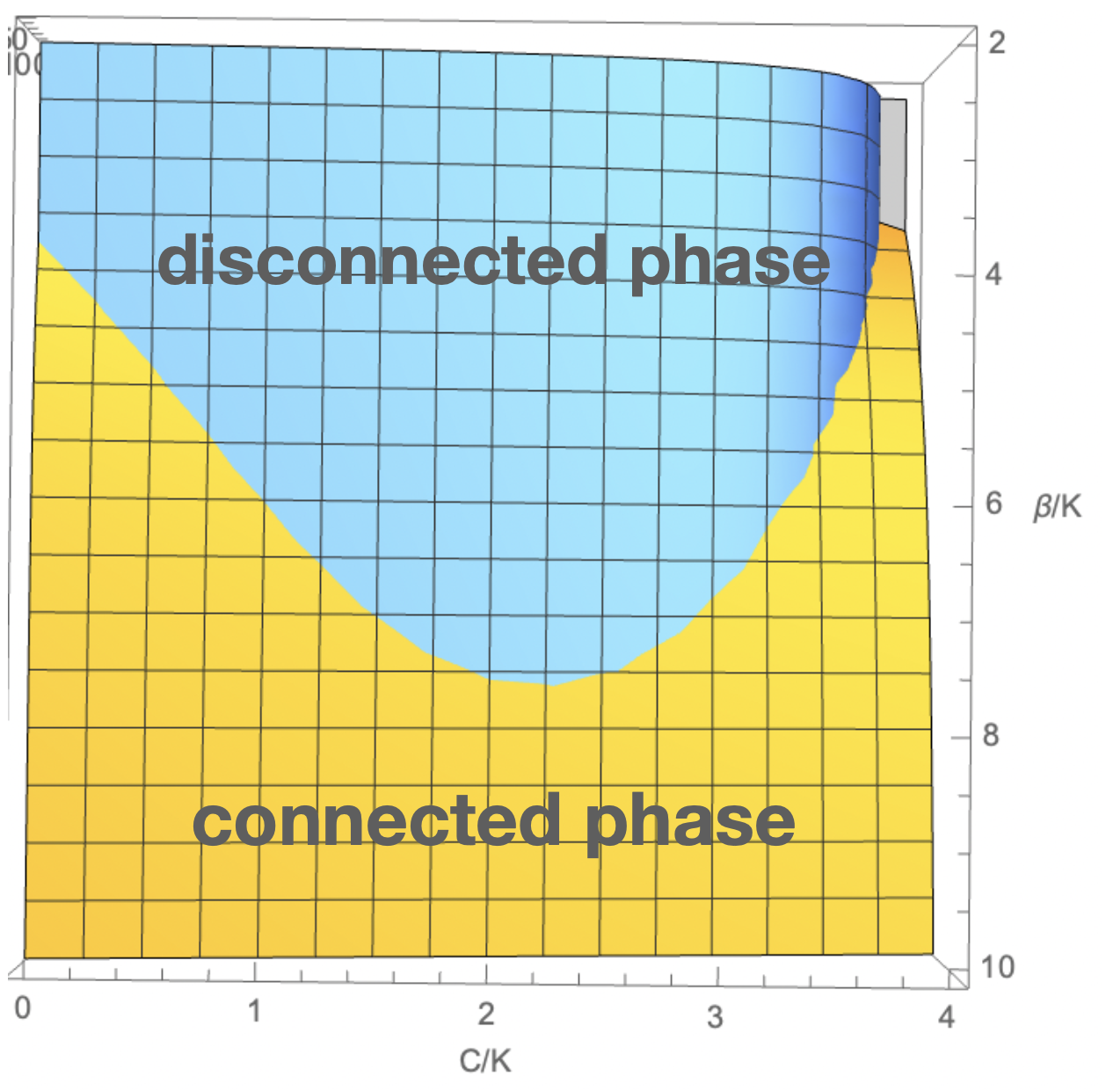}
        \caption{The phase structure on $\beta/K$ and $C/K$ with constant $\mathcal{E}$.}
        \label{fig:transition_CoverK}
      \end{center}
    \end{figure}

When $\beta/K$ is relatively small, there can exist one or two phase transitions on $C/K$. When $C/K=0$, depending on $\beta/K$ and $ \mathcal{E}$, we start with either the connected phase (relatively medium $\beta/K$) or the disconnected phase (relatively small $\beta/K$). As we increase $C/K$, the entropy approaches the disconnected phase. Hence, there is a phase transition on $C/K$ from the connected phase to the disconnected phase, with $C/K=0$ belonging to the connected phase. When further increasing $C/K$ to the upper bound $\mathcal{E}^2$, we will go back to the connected phase, since $S_0$ diverges and dominates when $C/K \to \mathcal{E}^2$ (see Fig.~\ref{fig:transition_CoverK}).

\textit{Quantum corrections.---} Adding quantum correction to the Schwarzian partition function would introduce additional logarithmic terms in $S^{\text{conn}}$ \cite{Saad:2019lba, Mertens:2022irh, Turiaci:2024cad}:
\be
S^{\text{conn}}_{\text{quantum}} = S^{\text{conn}} + \frac{3}{2} \log\frac{2\pi C}{\beta}+\frac{1}{2} \log\frac{2\pi C}{\beta}\, ,
\ee
where $\frac{3}{2} \log\frac{2\pi C}{\beta}$ comes from the disk quantum correction, and $\frac{1}{2} \log\frac{2\pi C}{\beta}$ comes from the trumpet correction.

In a special limit $K\ll C$, the ratio $\beta/K$ cannot increase all the way to $\infty$. Instead, it should be truncated at $\beta_q/K=\frac{2\pi C}{K}$. In this case, for the discussion in case i) on the transition of $\beta/K$, the sum $S^{\text{conn}} +2S_0 $ can only decrease to the minimal value at $\beta=\beta_q$ about
\be
(S^{\text{conn}} + 2S_0)_{\textrm{min}} \approx \left(\frac{8\pi^2 C}{\beta} + 2\log\frac{\beta_q}{\beta} \right)_{\beta = \beta_q} = 4\pi\, .
\ee
If additionally $4\pi > 2S_0$, the predicted transition point $\beta_c$ is larger than $\beta_q$ (or equivalently, $T_c < T_q$), the transition falls outside the working regime of consideration ($T>T_q$). In contrast, when $4\pi < 2S_0$, the quantum-corrected term $2\log\frac{\beta_q}{\beta}$ only shifts the transition point discussed in Case i), which considers phase transitions without quantum corrections.

\textit{Discussion.---} We considered the boundary effective theory of near-extremal AdS RN black holes in the near-horizon region. The theory consists of a Schwarzian mode and a $U(1)$ phase mode. We computed the partition function on replica geometries as $n\to 1$ and derived the entropy from the boundary perspective. Our analysis uncovered a rich phase structure, with transitions between connected and disconnected replica saddles controlled by the temperature $\beta$ and the couplings ($C$, $K$, $\mathcal{E}$) of the effective theory.

For future directions, the topological term in the effective theory \eqref{action_Sch_U(1)} exhibits a transition at $\mathcal{E}^2 = C/K$, where the topological entropy $S_0$ becomes imaginary. This suggests a possible unified boundary description for both dS$_2$ and AdS$_2$ JT gravity theories. Exploring this connection could shed light on the relation between quantum gravity in de Sitter and anti-de Sitter spacetimes from a boundary perspective.

\section*{acknowledgments}

We would like to thank Zi-Qing Xiao and Jinwu Ye for many helpful discussions. J.~N. was supported in part by the NSFC under grants No.~12375067 and No.~12547104.

\bibliography{ref}

\providecommand{\href}[2]{#2}\begingroup\raggedright\begin{thebibliography}{10}

\bibitem{Cadoni:2008mw}
M.~Cadoni and M.~R. Setare, ``{Near-horizon limit of the charged BTZ black hole
  and AdS(2) quantum gravity},''
  \href{http://dx.doi.org/10.1088/1126-6708/2008/07/131}{{\em JHEP} {\bfseries
  07} (2008) 131}, \href{http://arxiv.org/abs/0806.2754}{{\ttfamily
  arXiv:0806.2754 [hep-th]}}.

\bibitem{Almheiri:2016fws}
A.~Almheiri and B.~Kang, ``{Conformal Symmetry Breaking and Thermodynamics of
  Near-Extremal Black Holes},''
  \href{http://dx.doi.org/10.1007/JHEP10(2016)052}{{\em JHEP} {\bfseries 10}
  (2016) 052}, \href{http://arxiv.org/abs/1606.04108}{{\ttfamily
  arXiv:1606.04108 [hep-th]}}.

\bibitem{Moitra:2018jqs}
U.~Moitra, S.~P. Trivedi, and V.~Vishal, ``{Extremal and near-extremal black
  holes and near-CFT$_{1}$},''
  \href{http://dx.doi.org/10.1007/JHEP07(2019)055}{{\em JHEP} {\bfseries 07}
  (2019) 055}, \href{http://arxiv.org/abs/1808.08239}{{\ttfamily
  arXiv:1808.08239 [hep-th]}}.

\bibitem{Iliesiu:2019lfc}
L.~V. Iliesiu, ``{On 2D gauge theories in Jackiw-Teitelboim gravity},''
  \href{http://arxiv.org/abs/1909.05253}{{\ttfamily arXiv:1909.05253
  [hep-th]}}.

\bibitem{Iliesiu:2020qvm}
L.~V. Iliesiu and G.~J. Turiaci, ``{The statistical mechanics of near-extremal
  black holes},'' \href{http://dx.doi.org/10.1007/JHEP05(2021)145}{{\em JHEP}
  {\bfseries 05} (2021) 145}, \href{http://arxiv.org/abs/2003.02860}{{\ttfamily
  arXiv:2003.02860 [hep-th]}}.

\bibitem{Sachdev:2019bjn}
S.~Sachdev, ``{Universal low temperature theory of charged black holes with
  AdS$_2$ horizons},'' \href{http://dx.doi.org/10.1063/1.5092726}{{\em J. Math.
  Phys.} {\bfseries 60} no.~5, (2019) 052303},
  \href{http://arxiv.org/abs/1902.04078}{{\ttfamily arXiv:1902.04078
  [hep-th]}}.

\bibitem{Davison:2016ngz}
R.~A. Davison, W.~Fu, A.~Georges, Y.~Gu, K.~Jensen, and S.~Sachdev,
  ``{Thermoelectric transport in disordered metals without quasiparticles: The
  Sachdev-Ye-Kitaev models and holography},''
  \href{http://dx.doi.org/10.1103/PhysRevB.95.155131}{{\em Phys. Rev. B}
  {\bfseries 95} no.~15, (2017) 155131},
  \href{http://arxiv.org/abs/1612.00849}{{\ttfamily arXiv:1612.00849
  [cond-mat.str-el]}}.

\bibitem{Gaikwad:2018dfc}
A.~Gaikwad, L.~K. Joshi, G.~Mandal, and S.~R. Wadia, ``{Holographic dual to
  charged SYK from 3D Gravity and Chern-Simons},''
  \href{http://dx.doi.org/10.1007/JHEP02(2020)033}{{\em JHEP} {\bfseries 02}
  (2020) 033}, \href{http://arxiv.org/abs/1802.07746}{{\ttfamily
  arXiv:1802.07746 [hep-th]}}.

\bibitem{Penington:2019kki}
G.~Penington, S.~H. Shenker, D.~Stanford, and Z.~Yang, ``{Replica wormholes and
  the black hole interior},''
  \href{http://dx.doi.org/10.1007/JHEP03(2022)205}{{\em JHEP} {\bfseries 03}
  (2022) 205}, \href{http://arxiv.org/abs/1911.11977}{{\ttfamily
  arXiv:1911.11977 [hep-th]}}.

\bibitem{Almheiri:2019qdq}
A.~Almheiri, T.~Hartman, J.~Maldacena, E.~Shaghoulian, and A.~Tajdini,
  ``{Replica Wormholes and the Entropy of Hawking Radiation},''
  \href{http://dx.doi.org/10.1007/JHEP05(2020)013}{{\em JHEP} {\bfseries 05}
  (2020) 013}, \href{http://arxiv.org/abs/1911.12333}{{\ttfamily
  arXiv:1911.12333 [hep-th]}}.

\bibitem{Goto:2020wnk}
K.~Goto, T.~Hartman, and A.~Tajdini, ``{Replica wormholes for an evaporating 2D
  black hole},'' \href{http://dx.doi.org/10.1007/JHEP04(2021)289}{{\em JHEP}
  {\bfseries 04} (2021) 289}, \href{http://arxiv.org/abs/2011.09043}{{\ttfamily
  arXiv:2011.09043 [hep-th]}}.

\bibitem{Marolf:2020xie}
D.~Marolf and H.~Maxfield, ``{Transcending the ensemble: baby universes,
  spacetime wormholes, and the order and disorder of black hole information},''
  \href{http://dx.doi.org/10.1007/JHEP08(2020)044}{{\em JHEP} {\bfseries 08}
  (2020) 044}, \href{http://arxiv.org/abs/2002.08950}{{\ttfamily
  arXiv:2002.08950 [hep-th]}}.

\bibitem{Cotler:2019nbi}
J.~Cotler, K.~Jensen, and A.~Maloney, ``{Low-dimensional de Sitter quantum
  gravity},'' \href{http://dx.doi.org/10.1007/JHEP06(2020)048}{{\em JHEP}
  {\bfseries 06} (2020) 048}, \href{http://arxiv.org/abs/1905.03780}{{\ttfamily
  arXiv:1905.03780 [hep-th]}}.

\bibitem{Maldacena:2019cbz}
J.~Maldacena, G.~J. Turiaci, and Z.~Yang, ``{Two dimensional Nearly de Sitter
  gravity},'' \href{http://dx.doi.org/10.1007/JHEP01(2021)139}{{\em JHEP}
  {\bfseries 01} (2021) 139}, \href{http://arxiv.org/abs/1904.01911}{{\ttfamily
  arXiv:1904.01911 [hep-th]}}.

\bibitem{Mertens:2019tcm}
T.~G. Mertens and G.~J. Turiaci, ``{Defects in Jackiw-Teitelboim Quantum
  Gravity},'' \href{http://dx.doi.org/10.1007/JHEP08(2019)127}{{\em JHEP}
  {\bfseries 08} (2019) 127}, \href{http://arxiv.org/abs/1904.05228}{{\ttfamily
  arXiv:1904.05228 [hep-th]}}.

\bibitem{Cordes:1994fc}
S.~Cordes, G.~W. Moore, and S.~Ramgoolam, ``{Lectures on 2-d Yang-Mills theory,
  equivariant cohomology and topological field theories},''
  \href{http://dx.doi.org/10.1016/0920-5632(95)00434-B}{{\em Nucl. Phys. B
  Proc. Suppl.} {\bfseries 41} (1995) 184--244},
  \href{http://arxiv.org/abs/hep-th/9411210}{{\ttfamily arXiv:hep-th/9411210}}.

\bibitem{Blommaert:2018oue}
A.~Blommaert, T.~G. Mertens, and H.~Verschelde, ``{Edge dynamics from the path
  integral {\textemdash} Maxwell and Yang-Mills},''
  \href{http://dx.doi.org/10.1007/JHEP11(2018)080}{{\em JHEP} {\bfseries 11}
  (2018) 080}, \href{http://arxiv.org/abs/1804.07585}{{\ttfamily
  arXiv:1804.07585 [hep-th]}}.

\bibitem{Saad:2019lba}
P.~Saad, S.~H. Shenker, and D.~Stanford, ``{JT gravity as a matrix integral},''
  \href{http://arxiv.org/abs/1903.11115}{{\ttfamily arXiv:1903.11115
  [hep-th]}}.

\bibitem{Mertens:2022irh}
T.~G. Mertens and G.~J. Turiaci, ``{Solvable models of quantum black holes: a
  review on Jackiw{\textendash}Teitelboim gravity},''
  \href{http://dx.doi.org/10.1007/s41114-023-00046-1}{{\em Living Rev. Rel.}
  {\bfseries 26} no.~1, (2023) 4},
  \href{http://arxiv.org/abs/2210.10846}{{\ttfamily arXiv:2210.10846
  [hep-th]}}.

\bibitem{Turiaci:2024cad}
G.~J. Turiaci, ``{Les Houches lectures on two-dimensional gravity and
  holography},'' \href{http://arxiv.org/abs/2412.09537}{{\ttfamily
  arXiv:2412.09537 [hep-th]}}.

\end{thebibliography}\endgroup

\bibliographystyle{utphys}

\end{document}